\documentclass[12pt]{article}
\usepackage{epsf}
\usepackage{a4wide,graphicx}
\usepackage{cite}
\usepackage{amsmath}

\newcommand{\mev}{\textrm{ MeV}}
\newcommand{\MeV}{\textrm{ MeV}}
\newcommand{\Ld}{\Lambda}

\newcommand{\Sg}{\Sigma}
\newcommand{\Sgs}{\Sigma^*}

\newcommand{\bk}{\bar{K}}

\newcommand{\be}{\begin{equation}}
\newcommand{\ee}{\end{equation}}
\newcommand{\ba}{\begin{eqnarray}}
\newcommand{\ea}{\end{eqnarray}}
\newcommand{\Ls}{{\Lambda(1520)}}
\newcommand{\rw}{\rightarrow}
\newcommand{\pr}{^{\,\prime}}

\newcommand{\tph}{{\tilde \phi}}
\newcommand{\half}{\tfrac{1}{2}}
\newcommand{\thalf}{\tfrac{3}{2}}

\begin{document}

\title{Unitary coupled channel analysis of the $\Lambda(1520)$
resonance}

\author{
L.~Roca, Sourav~Sarkar, V.~K.~Magas and E.~Oset\\
{\small Departamento de F\'{\i}sica Te\'orica and IFIC,
Centro Mixto Universidad de Valencia-CSIC,} \\ 
{\small Institutos de
Investigaci\'on de Paterna, Aptdo. 22085, 46071 Valencia, Spain}\\ 
}

\date{\today}

\maketitle

\begin{abstract} 

We study the $\Lambda(1520)$ resonance in a coupled channel
approach involving the $\pi\Sigma(1385)$, $K\Xi(1530)$,
$\bar{K}N$ and $\pi\Sigma$ channels. Implementing unitarity in
coupled channels, we make an analysis of the relative importance
of the different mechanisms which contribute to the dynamical
structure of this resonance. From experimental information on
some partial wave amplitudes and constraints  imposed by
unitarity, we get a comprehensive description of the amplitudes
and hence the couplings to the different channels. We test these
amplitudes in different reactions like $K^-p\to\Lambda\pi\pi$,
$\gamma p\to K^+K^-p$, $\gamma p\to K^+\pi^0\pi^0\Lambda$ and
$\pi^- p\to K^0 K^-p$ and find a fair agreement with the
experimental data.

\end{abstract}

\section{Introduction}

The $\Ls$ ($3/2^-$, $D_{03}$) resonance is capturing renewed
attention, particularly since it appears invariably in searches
of pentaquarks in photononuclear reactions
\cite{nakano,Joo:2005gs} in $\gamma p\to K^+K^-p$ and $\gamma
d\to\ K^+K^-np$. Since getting signals for pentaquarks involves
cuts in the spectrum and subtraction of backgrounds, the
understanding of the resonance properties and the strength of
different reactions in the neighborhood of the peak becomes
important in view of a correct interpretation of invariant mass
spectra when cuts and background subtractions are made.

There is another reason that justifies a thorough study of the
$\Ls$ resonance. Indeed, much progress has been done interpreting
the low lying $1/2^-$ resonances as dynamically generated from
the interaction of the octet of pseudoscalar mesons with the
octet of stable baryons
\cite{Kaiser:1996js,Oset:1997it,Oller:2000fj,Garcia-Recio:2002td,Hyodo:2003jw}
which allows one to make predictions for resonance formation in
different reactions \cite{Oller:2000ma}. One of the surprises on
this issue was the realization that the $\Lambda(1405)$ resonance
is actually a superposition of two states, a wide one coupling
mostly to $\pi\Sigma$ and a narrow one coupling mostly to $\bar K
N$ \cite{Oller:2000fj,Jido:2003cb,Garcia-Recio:2003ks}. The
performance of a recent experiment on the
$K^-p\to\pi^0\pi^0\Sigma^0$ reaction \cite{Prakhov:2004an} and
comparison with older ones, particularly the $\pi^-p\to
K^0\pi\Sigma$ reaction \cite{Thomas:1973uh},
 has brought evidence
on the existence of these two $\Lambda(1405)$ states
\cite{Magas:2005vu}. Extension of these works to the interaction
of the octet of pseudoscalar mesons with the decuplet of baryons
has led to the conclusion that the low lying $3/2^-$ baryons are
mostly dynamically generated objects \cite{lutz,Sarkar:2004jh}.
One of these states is the $\Ls$ resonance which is generated
from the interaction of the coupled channels $\pi\Sigma(1385)$
and $K\Xi(1530)$, and couples mostly to the first channel to the
point that, in this picture, the state would qualify as a
quasibound $\pi\Sigma^*$ state. Indeed, the nominal mass of the
$\Ls$ is a few$\MeV$ below the average of the $\pi\Sigma^*$
mass. However, the PDG \cite{Eidelman:2004wy} gives a width of
$15\mev$ for the $\Ls$, with branching ratios of $45$\% 
into $\bar KN$ and $43$\%
 into $\pi\Sigma$, and only a small branching ratio
of the order of $4$\% 
for $\pi\Sigma^*$ which could be of the
order of $10$\%
 according to some analysis \cite{Mast:1973gb}
which claims that about $85$\% 
of the decay into $\pi\pi\Lambda$
is actually $\pi\Sigma^*$. The association of $\pi\pi\Lambda$ to
$\pi\Sigma^*$ in the peak of the $\Ls$ is a non trivial test
since one has no phase space for $\pi\Sigma^*$ excitation and
only the width of the $\Sigma^*$ allows for this decay, hence
precluding the reconstruction of the $\Sigma^*$ resonant
 shape from the
$\pi\Lambda$ decay product. Our theoretical study here
will allow a more precise determination from the study of
$K^-p\to\pi^0\pi^0\Lambda$ reaction \cite{Prakhov:2004ri}, which
proceeds mostly via
$K^-p\to\pi^0(\Sigma^{*0})\to\pi^0(\pi^0\Lambda)$ and involves
the $\Sigma^*$ propagator, overcoming the
reconstruction of the $\Sgs$ through the $\pi\Lambda$ invariant mass. In any case, the
large branching ratios to $\bar K N$ and $\pi \Sigma$, of the
order of $90$\% together, indicate that the $\bar K N$ and
$\pi\Sigma$ channels must play a relevant role in building up the
resonance.

In the present work we tackle this problem by performing a coupled
channel analysis of the $\Ls$ data with $\pi\Sigma^*$, $K\Xi^*$,
$\bar{K}N$ and $\pi\Sigma$, the first two channels interacting in
$s$-wave and the last two channels in $d$-wave to match 
the $3/2^-$ spin and
parity of the $\Ls$ resonance.

Anticipating results we shall see that although the
$\pi\Sigma^*$ remains with the largest coupling to $\Ls$, its
strength is reduced with respect to the simpler picture of only
$\pi\Sigma^*$ building up the resonance, and at the same time
there is a substantial coupling to $\bar KN$ and $\pi\Sigma$
which distorts the original $\pi\Sigma^*$ quasibound picture and
makes the $\bar KN$ and $\pi\Sigma$ channels relevant in the
interpretation of different reactions.

The procedure followed in this paper is the following: first we carry
out a fit to $\bar KN\to\bar KN$ and $\bar KN\to\pi\Sigma$ data in
$d$-waves for $I=0$ in order to determine a few unknown parameters
beyond the transition potentials 
of the $\pi\Sigma^*$, $K\Xi^*$, subsystems which are provided
by chiral Lagrangians \cite{Sarkar:2004jh,lutz}. With
this input we make predictions for the $K^-p\to\pi^0\pi^0\Lambda$ and
$K^-p\to\pi^+\pi^-\Lambda$ reactions, providing absolute cross
sections in good agreement with experiment. Furthermore, we also make
predictions for the shape of the $K^-p$ mass distribution for  the
$\gamma p\to K^+K^-p$ and $\pi^- p\to K^+K^-p$ reactions and for the
absolute value of the ratio of the $\gamma p\to
K^+\pi^0\pi^0\Lambda$ and $\gamma p\to K^+K^-p$ 
reactions at the peak of the
$\Ls$ resonance, for which there are no
 experimental data available.

The work provides a good model for the $\Ls$ resonance  within a
unitary coupled channel approach, which goes beyond the simpler
picture of the resonance as a quasibound $\pi\Sigma^*$ state, and
provides a framework to study other reactions involving the production
of the $\Ls$.
 
%%%%%%%%%%%%%%%%%%%%%%%%%%%%%%%%%%%%%%%%%%%%%%%%%%%%%%%%%%%%%%
%%%%%%%%%%%%%%%%%%%%%%%%%%%%%%%%%%%%%%%%%%%%%%%%%%%%%%%%%%%%%%
\section{Formalism}

In Ref.\cite{Sarkar:2005ap} the $\Lambda(1520)$ resonance was
studied within a coupled channel formalism including the
$\pi\Sigma^*$, $K\Xi^*$ in $s$-wave and the $\bar K N$  and
$\pi\Sigma$ in $d$-waves leading to a good reproduction of the
pole position of the $\Lambda(1520)$ of the scattering
amplitudes. However, the use of the pole position to get the
properties of the resonance is far from being accurate as soon
as a threshold is opened close to the pole position on the real
axis, which is the present case with the $\pi\Sigma^*$ channel.
Apart from that, in the approach of Ref.~\cite{Sarkar:2005ap}
some matrix elements in the kernel of the Bethe-Salpeter (BS)
equation were not considered. In the present work we aim at a
more  precise description of the physical processes involving
the $\Lambda(1520)$ resonance. Hence, we introduce the rest of
tree level transition potentials relevant for the analysis:
$\bar{K}N\to\bar{K}N$, $\bar{K}N\to\pi\Sigma$ and
$\pi\Sigma\to\pi\Sigma$. Analogously to
Ref.~\cite{Sarkar:2005ap}, we use for these vertices effective
transition potentials which are 
proportional to the incoming and outgoing momentum squared 
in order to account for the $d$-wave character of the
channels. Denoting  $\pi\Sigma^*$, $K\Xi^*$,  $\bar K N$  and
$\pi\Sigma$ channels by $1$, $2$, $3$ and $4$ respectively, the
matrix containing the tree level amplitudes is written as:

\renewcommand{\arraystretch}{1.25}
\be
V=\left| 
\begin{array}{cccc}
C_{11}(k_1^0+k_1^0)\ & C_{12}(k_1^0+k_2^0) & \gamma_{13}\,q_3^2 &\gamma_{14}\,q^2_4 \\
C_{21}(k_2^0+k_1^0)\ & C_{22}(k_2^0+k_2^0) & 0 & 0 \\
\gamma_{13}\,q_3^2 & 0 & \gamma_{33}\, q^4_3 & \gamma_{34} \,q_3^2 \,q^2_4\\
\gamma_{14}\,q^2_4  & 0 & \gamma_{34} \,q_3^2 \,q_4^2 &  \gamma_{44}\, q^4_4
\end{array}
\right|~,
\label{eq:Vmatrix}
\ee
\noindent
where $q_i=\frac{1}{2\sqrt{s}}\sqrt{[s-(M_i+m_i)^2][s-(M_i-m_i)^2]}$,
$k_i^0=\frac{s-M_i^2+m_i^2}{2\sqrt{s}}$ 
and $M_i(m_i)$ is the baryon(meson) mass. 
The coefficients $C_{ij}$ are $C_{11}=\frac{-1}{f^2}$,
 $C_{21}=C_{12}=\frac{\sqrt{6}}{4f^2}$ and $C_{22}=\frac{-3}{4f^2}$,
where $f$ is $1.15f_\pi$, with $f_\pi$ ($=93\mev$) the pion decay constant,
which is an average between $f_\pi$ and
  $f_K$ as was used in Ref.~\cite{Oset:1997it} in
  the related problem of the dynamical 
  generation of the $\Lambda(1405)$.

The elements $V_{11}$, $V_{12}$, $V_{21}$, $V_{22}$ come from the
lowest order chiral Lagrangian involving the decuplet of baryons and
the octet of pseudoscalar mesons as discussed in
 Ref.~\cite{Sarkar:2004jh,lutz}.
We neglect the elements $V_{23}$ and $V_{24}$ which involve the
tree level interaction of the $K\Xi^*$ channel to the $d$-wave channels
because the $K\Xi^*$ threshold is far away from the $\Ls$ and its role
in the resonance structure is far smaller than that of the
$\pi\Sigma^*$.

In the literature several unitarization procedures have been
used to obtain a scattering matrix fulfilling
unitarity in coupled channels, like the Inverse Amplitude
Method \cite{Dobado:1996ps,ollerIAM} or the $N/D$ method
\cite{ollerND}. In this latter work the equivalence with the
Bethe-Salpeter equation used in \cite{ollernpa} was established.

In the present work we use the Bethe-Salpeter
equation with the $V$ given above as the kernel to obtain the
unitarized amplitude $T$:

\be
T=V+VGT \Rightarrow T=[1-VG]^{-1}V
\label{eq:bethe}
\ee

\begin{figure}
\begin{center}
\includegraphics[width=0.85\textwidth]{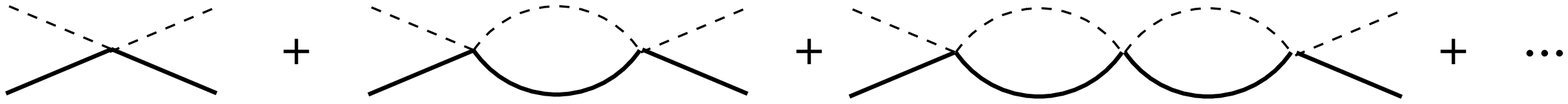}
\caption{Diagrammatic representation of the resummation of
meson-baryon loops
in the unitarization procedure.}
\label{fig:bethe}
\end{center}
\end{figure}

Diagrammatically this means that one is resumming the
series expressed in Fig.~\ref{fig:bethe}.
Following~\cite{Sarkar:2005ap}, the relationship of the
scattering matrix $T$ used in the unitary framework to the actual $t$
matrix in the Mandl-Shaw normalization~\cite{mandl} is given by
\ba
t_{\pi\Sgs\rw\pi\Sgs}&=&T_{\pi\Sgs\rw\pi\Sgs}\,, \nonumber\\
t_{K\Xi^*\rw K\Xi^*}&=&T_{K\Xi^*\rw K\Xi^*}\,, \nonumber\\
t_{\bk N\rw\pi\Sgs}&=&T_{\bk N\rw\pi\Sgs}\ {\cal C}(\half\ 2\
\thalf;m,M-m)Y_{2,m-M}(\hat{k})(-1)^{M-m}\sqrt{4\pi}\,, \nonumber\\
t_{\pi\Sg\rw\pi\Sgs}&=&T_{\pi\Sg\rw\pi\Sgs}\ {\cal C}(\half\ 2\
\thalf;m,M-m)Y_{2,m-M}(\hat{k})(-1)^{M-m}\sqrt{4\pi}\,, \nonumber\\
t_{\bk N\rw\bk N}&=&T_{\bk N\rw\bk N}\ \sum_M{\cal C}(\half\ 2\
\thalf;m,M-m)Y_{2,m-M}(\hat{k})\,\times \nonumber\\
&\times& {\cal C}(\half\ 2\
\thalf;m\pr,M-m\pr)Y_{2,m\pr-M}^*(\hat{k}\pr)(-1)^{m\pr-m}\ 4\pi\,.
\label{eq:tT}
\ea
The amplitudes 
$t_{\bar K N\to\pi\Sigma}$ and $t_{\pi\Sigma\to\pi\Sigma}$ have the
same form as $t_{\bar K N\to\bar K N}$ but changing  
$T_{\bar K N\to\bar K N}$ by 
$T_{\bar K N\to\pi\Sigma}$ and $T_{\pi\Sigma\to\pi\Sigma}$
respectively.

In Eq.~(\ref{eq:bethe}) $G$ stands for a diagonal matrix containing
the loop functions involving a baryon  and a meson \cite{Oller:2000fj}
and is given by:

\begin{eqnarray}
G_{l}(\sqrt{s},M_l,m_l)&=& i \, 2 M_l \int \frac{d^4 q}{(2 \pi)^4} \,
\frac{1}{(P-q)^2 - M_l^2 + i \epsilon} \, \frac{1}{q^2 - m^2_l + i
\epsilon}  \nonumber \\ 
\!\!\!\!\!\!\!\!\!\!&=& \frac{2 M_l}{16 \pi^2} \left\{ a_l(\mu) + \ln
\frac{M_l^2}{\mu^2} + \frac{m_l^2-M_l^2 + s}{2s} \ln \frac{m_l^2}{M_l^2}
-2i\pi \frac{q_l}{\sqrt{s}}
\right. \nonumber \\
 & & \!\!\!\!\!\!\!\!\!\! \phantom{\frac{2 M}{16 \pi^2}} +
\frac{q_l}{\sqrt{s}}
\left[
\ln(s-(M_l^2-m_l^2)+2 q_l\sqrt{s})+
\ln(s+(M_l^2-m_l^2)+2 q_l\sqrt{s}) \right. \nonumber  \\
& &\!\!\!\!\!\!\!\!\!\! \left. \phantom{\frac{2 M}{16 \pi^2} +
\frac{q_l}{\sqrt{s}}}
\left. \hspace*{-1.2cm}- \ln(s-(M_l^2-m_l^2)-2 q_l\sqrt{s})-
\ln(s+(M_l^2-m_l^2)-2 q_l\sqrt{s}) \right]
\right\}\,,
\label{propdr}
\end{eqnarray}
\noindent
where 
$\mu$ is the scale of dimensional regularization, $s=P^2$ with $P$ the
total four-momentum of the meson-baryon system and
 $a_l$ are unknown
subtraction constants. For the loops related to the $s$-wave channels
($\pi\Sigma^*$ and $K\Xi^*$) it has a natural size of around $-2$,
as has been obtained in many different works 
\cite{Oller:2000fj,Ramos:2002xh,Oset:2001cn}. For the
$d$-wave channels  ($\bar K N$  and $\pi\Sigma$) there is no such
estimate in the literature. We expect it to be larger compared to
the $s$-wave subtraction constant since it is likely to contain part
of the off-shell  contribution of the potentials in the loop, which
are expected to be large for the $d$-wave vertices.
In Eq.~(\ref{eq:bethe}) the momentum dependence of the
tree level amplitudes has been factorized out of the loop integral.
For the $s$-wave vertices this has been justified {\it e.g.} 
in \cite{ollernpa,Oset:1997it}. For the
$d$-wave vertices we assume that the off-shell momentum dependences
can be reabsorbed in the couplings and the subtraction constants
which are free parameters in our scheme. We will discuss further on this
issue below.

Since the $\pi\Sigma^*$ threshold lies in the $\Lambda(1520)$
region the consideration of the width of the $\Sigma^*$ resonance
in the loop function is crucial in order to account properly for
this channel. The  $\Sigma^*$  width is about $35\MeV$ while the
$\Lambda(1520)$ width is only about $15\MeV$. Since the threshold
effects are very important in the  description involving coupled
channels (due {\it e.g.} to the opening of sources of imaginary
parts), this implies that the proper consideration of the 
spectral distribution of the  $\Sigma^*$ resonance is essential.
This is achieved through the convolution of the $\pi\Sigma^*$ loop
function with the spectral distribution considering the $\Sigma^*$
width:

\be
{G}_{\pi\Sigma^*}(\sqrt{s},M_{\Sigma^*},m_\pi)
\to\int_{M_{\Sigma^*}-2\Gamma_0}
^{M_{\Sigma^*}+2\Gamma_0}d\sqrt{s'}\ \frac{-1}{\pi}\,
\textrm{Im}\left[
\frac{1}{\sqrt{s'}-M_{\Sigma^*}+i\Gamma_{\Sigma^*}(s')/2}\right]
G_{\pi\Sigma^*}(\sqrt{s},\sqrt{s'},m_\pi)
\ee
\noindent
where
\ba \nonumber
\Gamma_{\Sigma^*}(s')=\Gamma_0
&&\left(0.88\frac{q^3(s',M_\Lambda^2,m_\pi^2)}
{q^3(M_\Sigma^{*2},M_\Lambda^2,m_\pi^2)}
\Theta(\sqrt{s'}-M_\Lambda-m_\pi)\right.\\
&+&\left. \quad0.12\frac{q^3(s',M_\Sigma^2,m_\pi^2)}
{q^3(M_\Sigma^{*2},M_\Sigma^2,m_\pi^2)}
\Theta(\sqrt{s'}-M_\Sigma-m_\pi)\right),
\label{eq:gammas}
\ea
\noindent
with $\Gamma_0$ being the on-shell $\Sigma^*$ total decay width.
In Eq.~(\ref{eq:gammas}) we have assumed a $p$-wave decay of the
$\Sigma^*$ into $\Lambda\pi$ ($88$\%) and  $\Sigma\pi$ ($12$\%).
This consideration of the $\SgsÂ$ width is also an improvement from the previous work of
Ref.~\cite{Sarkar:2005ap}.\\

Note that we could think of other states in our coupled channels of the
vector-meson--baryon (VB) type like $K^*N$, $\rho\Sigma$, etc. In fact, as we
discuss in detail after Eq.~(\ref{eq:tphoto1}), the $K^*N\Lambda(1520)$
coupling can be large in some models \cite{Sibirtsev:2005ns,Titov:2005kf}. The
explicit consideration of these channels in the coupled channel formalism is
unnecessary for the following reasons: the thresholds for the VB states quoted
above are $1830\MeV$ and $1967\MeV$, more than $300\MeV$ above the  energy of
the $\Lambda(1520)$. As a consequence, their contribution to the scattering
amplitudes around the $\Lambda(1520)$ region is very weakly energy dependent,
since the corresponding VB loop functions are weakly energy dependent far away
from the VB threshold. Thus, their contribution can be easily incorporated in
terms of the subtraction constants introduced in Eq.~(\ref{propdr}) which are
fitted to data. 
 This appears to be also the case in the study of the 
$\pi N$ interaction in \cite{mores}, where, for instance, $\rho$ exchange in
the t-channel (in tensor form) is used to generate higher order terms in the
framework, but not as $\rho N$ or $\rho \Delta$ ~s-channel intermediate states.
Note that, even if the contribution of these new channels was
not numerically negligible, this would not mean that such channels are 
important in the structure of the resonance, since what matters for the wave
function components are the derivatives of the resonance selfenergy from these
channels with respect to the energy \cite{fetter}.\\

As explained before, in Eq.~(\ref{eq:Vmatrix}) the $C_{ij}$ coefficients are
obtained from the lowest order chiral Lagrangians accounting for the
Weinberg-Tomozawa term. One could include in the kernel of the BS equation
contributions of the higher order Lagrangians of the pseudoscalar octet and
baryon decuplet interaction in the $SU(3)$ sector, but this has not been
thoroughly studied. Some work is however done in the $SU(2)$ sector
\cite{Fettes:2000bb}.  The situation is different  in the interaction of the
pseudoscalar octet with the baryon octet, where work has been done including
higher order Lagrangians \cite{Borasoy:2005ie,Oller:2005ig} in the strange
sector. However, it is
worth stressing that already a good reproduction of the data is obtained with
the lowest order chiral Lagrangian
\cite{Borasoy:2005ie,Oller:2005ig,Oset:1997it,Oller:2000fj}. The effect of 
higher order
Lagrangians can be accounted for to some extent by means of the subtraction
constants of the loop functions (or equivalently fixing a cutoff to data
\cite{Oset:1997it}). In the present work we have more subtraction constants, as
well as unknown $\gamma$ parameters, which are fitted to the data, so there is
plenty of room to effectively account for the effect of higher order
Lagrangians by means of all these free parameters. 

In the SU(2) sector there is more work.  Interesting developments are done in
Ref.~ \cite{bernard} dealing with pions and nucleons, where higher order
Lagrangians are introduced, together with the $\Delta(1232)$ as an explicit 
degree of freedom, in such a way as to respect decoupling. Decoupling 
\cite{decoupling} states that in 
the chiral limit the leading non analytic corrections (LNAC) to S-matrix
elements and related magnitudes are given solely
in terms of the meson and baryon degrees of freedom, U and B terms of the
Lagrangian, and that the inclusion of mesonic ($\rho, \omega$) or baryonic
($\Delta, N^*)$ resonances in the pertinent loops does not modify the LNACs.  
  Another interesting development is
done in \cite{mores} where the higher order terms are obtained by using
explicitly the exchange of resonances. The framework respects unitarity in
coupled channels and matches at low energies with the perturbative results of
\cite{fettes1}, thus providing an  example of the resonance saturation
hypothesis \cite{derafael} in the baryon sector. 
   Developments along these lines in the meson-octet baryon-decuplet interaction
    would be certainly welcome. 

\section{Results}

In the model described so far we have as unknown parameters
$\gamma_{13}$, $\gamma_{14}$, $\gamma_{33}$, $\gamma_{34}$,
$\gamma_{44}$  in the $V$ matrix. Apart from these, there is also
the freedom in the value of the subtraction constants in the loop
functions. We will consider one subtraction constant for the
$s$-wave channels ($a_0$) and one for the $d$-wave ones ($a_2$).
Despite the apparent large number of free parameters in the $V$
matrix, it is worth emphasizing that the largest matrix elements
are $V_{11}$, $V_{12}$ and $V_{22}$ \cite{Sarkar:2004jh}
which come from a chiral  Lagrangian \cite{manohar} without any
free parameters. Due to the $d$-wave behavior
the other ones are expected to
be smaller,  as we will see below.

In order to obtain these parameters we fit our model to the
experimental results on the $\bar K N$ and $\pi\Sigma$ scattering
amplitudes in $d$-wave and $I=0$.
We use experimental data from 
Refs.~\cite{Gopal:1976gs,Alston-Garnjost:1977rs} where 
$\bar K N\to\bar K N$ and $\bar K N\to\pi\Sigma$ amplitudes are
provided from partial wave analysis.
These experimental amplitudes are related to the
amplitudes of Eq.~(\ref{eq:bethe}) through 
\be
\tilde{T}_{ij}(\sqrt{s})=-\sqrt{\frac{M_iq_i}{4\pi\sqrt{s}}}
\sqrt{\frac{M_jq_j}{4\pi\sqrt{s}}}\,T_{ij}(\sqrt{s})
\label{eq:Ttilde}
\ee
where $M$ and $q$ are the baryon mass and the on-shell 
C.M. momentum of
the specific channel.

It is also interesting to make connection with another standard
notation for the amplitudes for the scattering of spin $0$
with spin $1/2$ particles:
\ba
M&=&f+ig\,\vec{\sigma}\cdot\hat{n}\nonumber\\
f&=&\frac{1}{q}\sum_{l=0}^{\infty}\left[(l+1)f_{l+}+lf_{l-}\right]
P_l (\cos\theta)\nonumber\\
g&=&\frac{1}{q}\sum_{l=1}^{\infty}\left[f_{l+}-f_{l-}\right]P'_l (\cos\theta)
\sin\theta,
\ea
\noindent
where $\hat{n}=(\vec{q}_i\times\vec{q}_j)/
|\vec{q}_i\times\vec{q}_j|$.
For $l=2$, $J=3/2$, only $f_{l-}$ contributes and we
find, given the normalizations introduced in Eq.~(\ref{eq:tT}),
\be
f_{l-}=\tilde T_{ii}
\ee
\noindent
and similarly for the amplitudes $\tilde T_{ij}$ where $i$, $j$ stand for 
 the $\bar K N$ and $\pi \Sigma$ channels. The
elastic cross section is 
given by $\sigma=4\pi l|f_{l-}|^2/q^2$ $(l=2)$.

We now write the
amplitude $T_{ij}$ close to a resonance peak as 
\be
T_{ij}=\frac{g_ig_j}{\sqrt{s}-M_R+i\Gamma/2},
\label{eq:TBW}
\ee
 (note that, for $l=2$, the $q_i^2$ factor is incorporated in
 $g_i$).
We then have
\be
\Gamma_i=\frac{g_i^2}{2\pi}\frac{M_i}{M_R}q_i
=-\frac{1}{2\pi}\frac{\Gamma}{2} Im T_{ii}\frac{M_i}{M_R}q_i
\ee
and hence
\be
B_i=\frac{\Gamma_i}{\Gamma}=Im \tilde{T}_{ii}(\sqrt{s}=M_R).
\ee
Note that due to the appearance of the $q_i$, $q_j$ factors, 
Eq.~(\ref{eq:Ttilde}) can only be applied for channels which are open. For those which
are close to threshold the decay can only proceed via the overlap  of the mass
distributions  of the resulting products including their width, with the mass
of the decaying resonance, and this  situation  requires another treatment, as
we shall see below.

In Fig.~\ref{fig:fit} we show the result of our fit to the experimental
data. 
\begin{figure}
\centerline{
\includegraphics[width=0.8\textwidth]{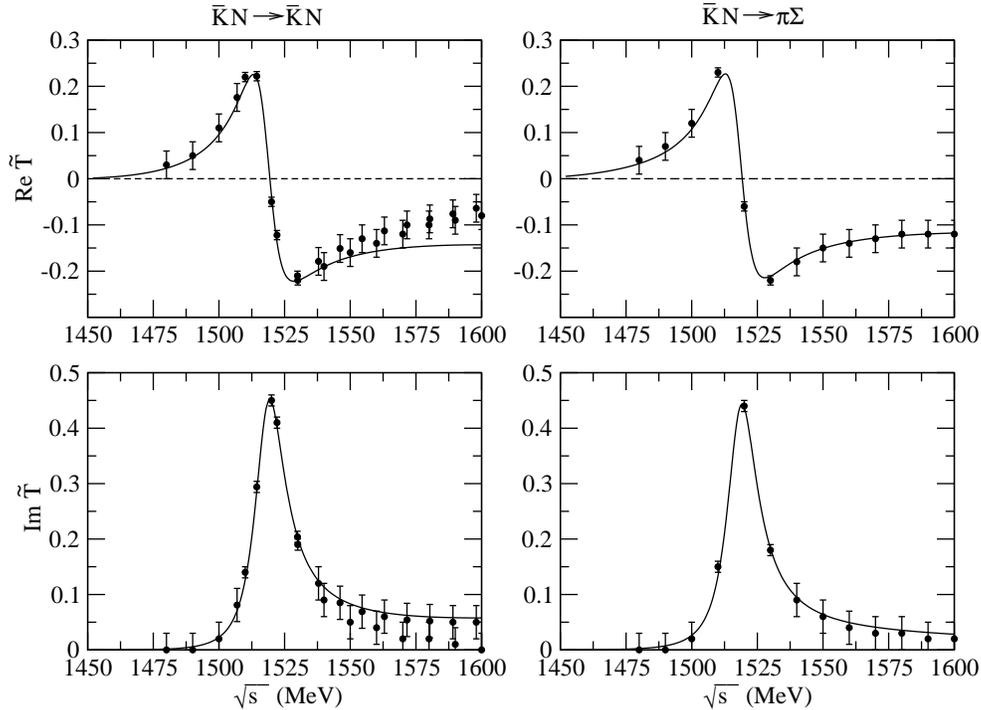}
}
\caption{Fit to the experimental amplitudes. 
Left column: $\bar K N\to\bar K N$; 
right column: $\bar K N\to\pi\Sigma$.  }
\label{fig:fit}
\end{figure}
The first column represents the real and imaginary parts (top and
bottom respectively) of the $\tilde{T}_{33}$ and the last column
denotes the same for $\tilde{T}_{34}$ along with the experimental
data of \cite{Gopal:1976gs,Alston-Garnjost:1977rs}.
In order to restrict the freedom in the parameters of the model we
have also introduced data for the $\pi\Sigma\to\pi\Sigma$ amplitude.
The data are not given in 
\cite{Gopal:1976gs,Alston-Garnjost:1977rs}. However, the results given in that
paper for the $\bar KN\to\bar KN$ and $\bar KN\to\pi\Sigma$ are results
of analysis of raw data. The same analysis would have provided
\be
T_{\pi\Sigma\to\pi\Sigma}=\frac{(T_{\bar KN\to\pi\Sigma})^2}
{T_{\bar KN\to\bar KN}}
\ee
given the resonant structure of the amplitudes. By introducing
these data we are forcing the resulting model to fulfill this
property, which helps constrain the freedom in the fit
parameters. In the data
from~\cite{Gopal:1976gs,Alston-Garnjost:1977rs} which we have 
considered, no errors are given.
 In order to perform a fit to the data some errors need to be
assigned. We have taken a reasonable criteria by assigning  an
error of $0.03$ to each point except the ones close to the peak
where an error of $0.01$ is taken to enforce the resonance
character of the data. We have checked that other reasonable
assumptions also lead to about the same solutions.
 It is worth noting that the shape of the amplitudes is rather
asymmetric, in the sense that it differs from a Breit-Wigner
shape. This is a consequence of the $d$-wave behavior and also  
of the non trivial internal dynamics imposed by unitarity in
coupled channels.

The values of the unknown parameters obtained from the fit 
are given in Table~\ref{tab:fit}.
\begin{table}
\begin{center}
\begin{tabular}{|c|c|c|c|c|c|c|}
\hline
  $a_0$ & $a_2$ & $\gamma_{13}$ $(\textrm{MeV}^{-3})$ & $\gamma_{14}$ $(\textrm{MeV}^{-3})$&
    $\gamma_{33}$ $(\textrm{MeV}^{-5})$&  $\gamma_{44}$
    $(\textrm{MeV}^{-5})$ &  $\gamma_{34}$ $(\textrm{MeV}^{-5})$ \\	
\hline
$-1.78$ & $-8.13$ & $0.98\times 10^{-7}$ & $1.10\times 10^{-7}$ & $-1.73\times 10^{-12}$
 & $-0.730\times 10^{-12}$ & $-1.108\times 10^{-12}$ \\
 $\pm 0.02$ & $\pm 0.03$ & $\pm 0.04\times 10^{-7}$ & $\pm 0.04\times 10^{-7}$ & 
 $\pm 0.02\times 10^{-12}$
 & $\pm 0.016\times 10^{-12}$ & $\pm 0.010\times 10^{-12}$ \\
\hline
\end{tabular}
\caption{Parameters obtained from fit}
\label{tab:fit}
\end{center}
\end{table}
We can see that the value obtained for the subtraction constant for
the $s$-wave channels ($a_0$) is of  natural size ($\sim-2$). 
Actually, the value of the $s$-wave loop 
function obtained using $a_0=-1.8$ agrees with the result obtained
with the cutoff method using a cutoff of about $500\mev$
(at
$\sqrt{s}\simeq 1520\mev$). On the other hand, regarding the $d$-wave loops,
the large value obtained for $a_2$ can be understood comparing also
to the cutoff method. If one keeps the momentum dependence of the
 $d$-wave vertices inside the
loop integral (i.e., one does not use the on-shell approximation
mentioned above) and evaluates the integral with the cutoff method,
then also a cutoff of about  $500\mev$  gives the
same result as the dimensional regularization with on-shell
factorization and $a_2\sim-8$.
In summary, the use of the dimensional regularization method
along with the on-shell factorization for both the $s$ and $d$-wave
loops, correspond to the result obtained with the cutoff method
without on-shell factorization using the same cutoff of about
$500\mev$. 

Next we make an estimate of the theoretical errors in the fit
parameters. We vary each parameter keeping the rest fixed such
as to increase the $\chi^2$ function by eight
units\footnote{This
is the equivalent in the case of seven parameters to changing the
$\chi^2$ by a unity when one has only one free parameter, which
is the standard procedure to get $68$\% confidence level
\cite{guardiola}.}
(out of the $110$ in the minimum). This leads to the errors
shown in  Table~\ref{tab:fit}. We shall see later on
 what repercussion they have
on the values of the coupling of the $\Lambda(1520)$
resonance to different channels. We can see that the errors are
small but they are enforced by the errors that we have assigned
to the data for the fit. However, this is consistent with
necessary small errors in the $g_3$ and $g_4$ couplings that we
shall see later on, since these couplings are related to the
$\bar{K}N$ and $\pi\Sigma$ branching ratios which are known within
about $1-2$\% precision.

The uncertainties in the parameters are larger if simultaneously we fit the
 other parameters to get a best fit to the data.  This means that there are
 correlations between the parameters. However, since what matters in the end is
 the errors in the couplings of the resonance to the different channels, which
 are the relevant physical quantities associated to the resonance, we can make 
 an alternative analysis of the errors by allowing each parameter in Table 1
  to change by
 a certain quantity and fitting simultaneously the other parameters to get a
 best fit to the data, with the $\chi^2$ increased by the same magnitude as
 before. At the end we have seven sets of parameters which allow us to get the
 couplings, $g_i$, with a certain dispersion.  The central values and
 uncertainties agree with those obtained by the former method.

\begin{figure}
\centerline{
\includegraphics[width=0.7\textwidth]{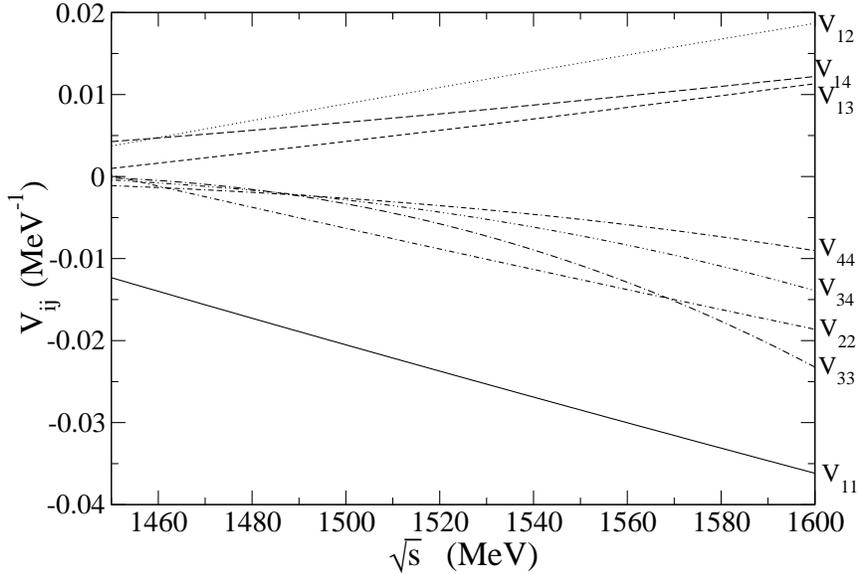}
}
\caption{Tree level potentials.}
\label{fig:V}
\end{figure}

In order to check that the values of the parameters 
$\gamma_{ij}$ obtained from the fit, 
support our earlier expectation regarding the dominance of the
chiral matrix elements as far as the $V$-matrix is concerned, we plot
in Fig.~\ref{fig:V} the different matrix elements as a function of
$\sqrt{s}$.
We can see that the largest elements are $V_{11}$ and $V_{12}$
in the region close to $1520\mev$. However, the loops play a
different role depending on the amplitudes. In the $\bar K
N\to\bar K N$ amplitude near $\sqrt{s}=1520\mev$, at one loop
level the $\bar K N$ and $\pi \Sigma$ loops largely dominate the
contribution. For the $\pi\Sgs\to\pi\Sgs$ amplitude at one loop
level, the $\pi\Sgs$ and $\pi\Sigma$ loops are the dominant ones
with similar strength. One should note that the tree level
amplitude in this latter 
case still exceeds the contribution of any of
the loops.

\begin{figure}
\centerline{
\includegraphics[width=0.9\textwidth,angle=0]{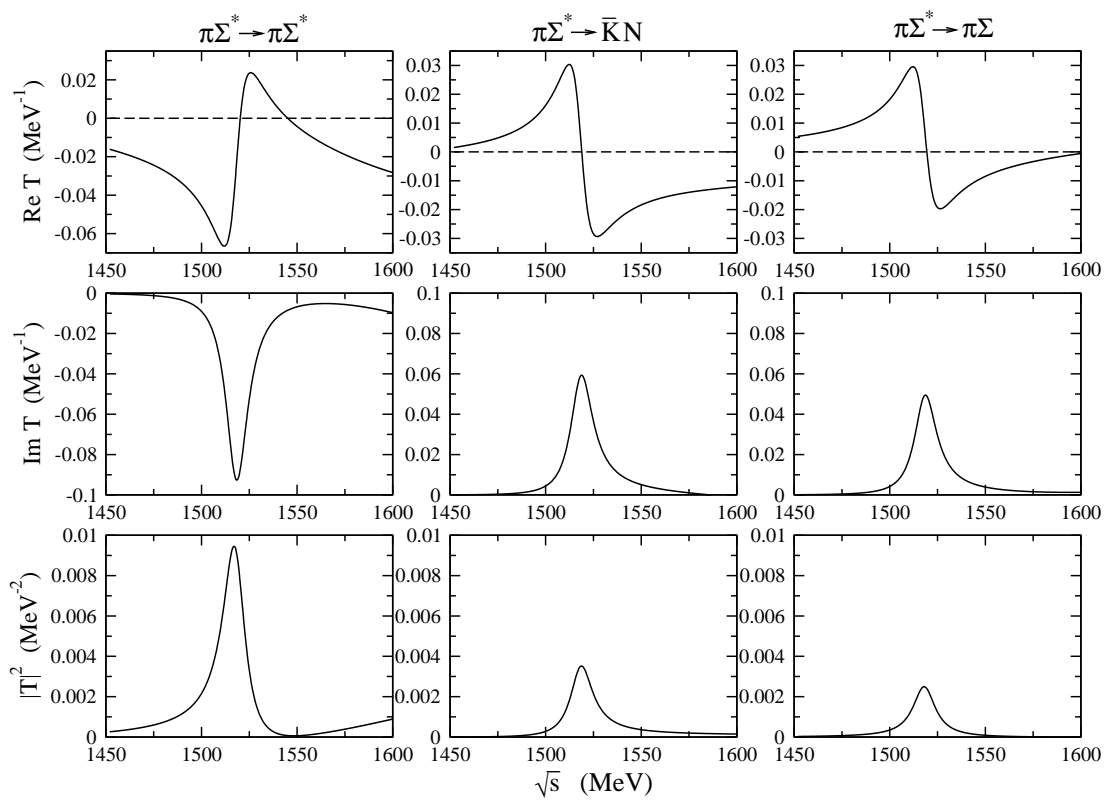}
}
\caption{Unitary amplitudes involving the $\pi\Sigma^*$ channel.
 From left to right:
 $\pi\Sigma^*\to\pi\Sigma^*$, $\pi\Sigma^*\to \bar K N$ and 
$\pi\Sigma^*\to \pi\Sigma$.}
\label{fig:Tij}
\end{figure}

Next we plot, in Fig.~\ref{fig:Tij}, the prediction for the
unitarized
amplitudes  for the different channels involving the $\pi\Sigma^*$.
From left to right the columns represent the
 $\pi\Sigma^*\to\pi\Sigma^*$, $\pi\Sigma^*\to \bar K N$ and 
$\pi\Sigma^*\to \pi\Sigma$ channels.
The rows denote from top to bottom the real part, imaginary part and modulus
squared of the amplitudes ($T_{ij}$) respectively.
We do not show the $K\Xi^*$ channel since it is less relevant as
an external state in physical processes. 

It is worth mentioning that the unitarization procedure does not only
provide the amplitude at the peak of the resonance but also 
obtains
the full amplitude at energies away from it. This can have
repercussion in some observables in specific physical processes as
we will discuss below.

From the imaginary part of the amplitudes it is straightforward to
obtain the couplings of the $\Ls$ to the different channels
in the following way. 
Close to the peak the amplitudes can be approximated 
by Eq.~(\ref{eq:TBW}),
which in this case reads
\be
T_{ij}(\sqrt{s})=\frac{g_i g_j}{\sqrt{s}-M_\Ls + i \Gamma_\Ls /2}
\label{eq:Tcoup}
\ee 
from where we have
\be
g_i g_j=-\frac{\Gamma_\Ls}{2} \frac{|T_{ij}(M_\Ls)|^2}
{Im[T_{ij}(M_\Ls)]},
\label{eq:gigj}
\ee
\noindent 
where $M_\Ls$ is the position of the peak in $|T_{ij}|^2$ and 
$\Gamma_\Ls=15.6\mev$.

Up to a global sign of one of the couplings (we choose $g_1$ to be
positive), the couplings we obtain are shown in Table~\ref{tab:coup}.
\begin{table}
\begin{center}
\begin{tabular}{|c|c|c|c|}
\hline
$g_1$ &$g_2$ &$g_3$ &$g_4$ \\	
\hline
 $0.91$ & $-0.29$ & $-0.54$ & $-0.45$ \\
  $\pm 0.06$ & $\pm 0.06$ & $\pm 0.01$ & $\pm 0.01$ \\
\hline
\end{tabular}
\caption{Couplings of the $\Ls$ resonance to the different channels}
\label{tab:coup}
\end{center}
\end{table}
We can see from the values that the $\Ls$ resonance 
couples most strongly to the $\pi\Sgs$ channel.
The fact that we are able to predict the value of this coupling is a
non trivial consequence of the unitarization procedure that we
employ.

In order to determine the errors in the $g_i$ couplings, we
propagate the errors of Table~\ref{tab:fit} into
Eq.~(\ref{eq:gigj}). Another source of uncertainty arises from 
the fact that in the $\pi\Sigma^*$ and $K\Xi^*$ channels the maximum
of $|T|^2$ and $Im T$ appear at slightly different energies
(about $4\MeV$). This induces also some uncertainty in the
evaluation of  Eq.~(\ref{eq:gigj}) and is also taken into
account in the errors shown in Table~\ref{tab:coup}.

\begin{figure}
\centerline{
\includegraphics[width=0.4\textwidth,angle=0]{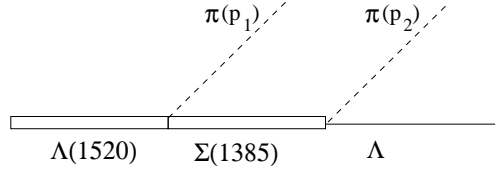}
}
\caption{Diagram for $\Ls$ decay into $\pi\pi\Lambda$}
\label{fig:Lambda_decay}
\end{figure}

With the value for $g_1$ obtained above,
 we  now evaluate the partial
decay width of the $\Lambda(1520)$ into $\pi\pi\Lambda$ 
assuming that this process is dominated by the $\pi\Sigma^*$ channel.
The diagram describing this decay is shown in 
Fig.~\ref{fig:Lambda_decay}.

The amplitude for the decay can be written as

\be
t(p_1,p_2)=
\frac{g_1}{\sqrt{s_{\Sgs}}-M_{\Sgs}+i\Gamma_{\Sgs}(\sqrt{s_{\Sgs}})/2}
\frac{f_{\Sgs\pi\Lambda}}{m_\pi}
\langle \frac{1}{2}m|\vec S \cdot \vec{p_2}|\frac{3}{2}M\rangle,
\ee

where $\vec p_2$ is taken in the rest frame of the $\Sgs$. The width is then given by

\ba
\Gamma_{\Ls\to\pi\pi\Lambda}=\frac{1}{2}\frac{1}{(2\pi)^4}
\frac{M_\Lambda M_{\Ls}}{4\sqrt{s}}
\int_{-1}^{1} & & d\cos\theta_1
 \int_{m_\pi}^{\omega_{max}}d\omega_1\ 
 \int_0^{2\pi}d\tph_2\ 
 \int_{m_\pi}^{\omega_{max}}d\omega_2  \cdot\nonumber \\
&\cdot& \frac{1}{4}\sum_{m,M}|t(p_1,p_2)+t(p_2,p_1)|^2
\ \Theta(1-|A|^2)
\label{eq:width}
\ea
where $A=\cos\theta_{12}=\frac{\left[(\sqrt{s}-\omega_1-\omega_2)^2
-|\vec p_1|^2-|\vec p_2|^2-M_\Ld^2\right]}{2\ |\vec p_1||\vec p_2|}
$ and $\omega_{max}=\frac{s+m_\pi^2-(m_\pi+M_\Ld)^2}{2\sqrt{s}}$.

With the coupling $g_1$ of Table~\ref{tab:coup},  and dividing the result of 
Eq.~(\ref{eq:width}) by 0.88 
(the branching ratio of the $\Sigma(1385)$ decay
into $\pi \Lambda$), we obtain a
branching ratio for $\Lambda(1520) \to  \pi \Sigma^*$ of around 0.14. 
By using the expression
$Im \tilde{T}_{ij}=\sqrt{B_i B_j}$ we can see from Fig.~\ref{fig:fit}
that the branching ratios for $\bar K N$
and $\pi\Sigma$ are $0.45$ and $0.43$ respectively. 
All these branching ratios
 essentially
sum up to unity considering the uncertainties in the calculations,
(in particular the exact position of the peak where the couplings are
evaluated). The branching ratio to $\pi \Sigma^*$ is small because of lack of phase space for the decay. However, the relevant magnitude concerning the nature of a
resonance is the coupling of the resonance to the different states. In this
sense the $g_1$ coupling is still the largest. Originally, we had a theory with
chiral Lagrangians which  provided  the $\Lambda(1520)$ as a dynamically generated
resonance from $\pi\Sgs$. In reality also the
 $\bar K N$ and $\pi\Sigma$ channels are present,
which are quite relevant and distort that approximate picture.
As a consequence, $g_1$ becomes reasonably smaller and
simultaneously one gets a relatively large $\bar K N$ and
$\pi\Sigma$ coupling. Yet, with this admittedly large distortion
of the original picture of the $\Lambda(1520)$ as a quasibound
$\pi\Sigma^*$ state, the physical $\Lambda(1520)$ seems to keep
a memory of this original picture which shows up in the coupling
$g_1$, such that $g_1^2$ is $2.8$
 times larger than the coupling
squared for the $\bar K N$ state.  This of course should
manifest itself in processes where the $\pi\Sigma^*$ channel
appears without restrictions of phase space, like in
$\Lambda(1520)$ decay in nuclei in that channel with the pion
becoming a $ph$ excitation.

The prediction of the amplitudes involving $\pi\Sigma^*$ channels can
be checked in particular reactions where this channel could play an
important role. First we evaluate the cross section for $K^-p\to\pi\pi\Lambda$ in the
lines of Ref.~\cite{Sarkar:2005ap} but using the new coupled channel
formalism. The mechanisms and the 
expressions
 for the amplitudes and
the cross sections can be found in Ref.~\cite{Sarkar:2005ap} where,
apart from the coupled channel unitarized amplitude, other
mechanisms\footnote{In the background processes
of Ref.~\cite{Sarkar:2005ap} we have now 
included the recoil corrections in the baryon-baryon-pseudoscalar
vertices. We substitute $\vec p_1$ in the curly brackets of Eqs.~(36) and
(37) of Ref.~\cite{Sarkar:2005ap} by $\vec p_1(1+p^0/2M)+\vec 
k \,p_1^0/M$.}
 of relevance above the $\Ls$ peak were included.

\begin{figure}
\centerline{
\includegraphics[width=0.6\textwidth]{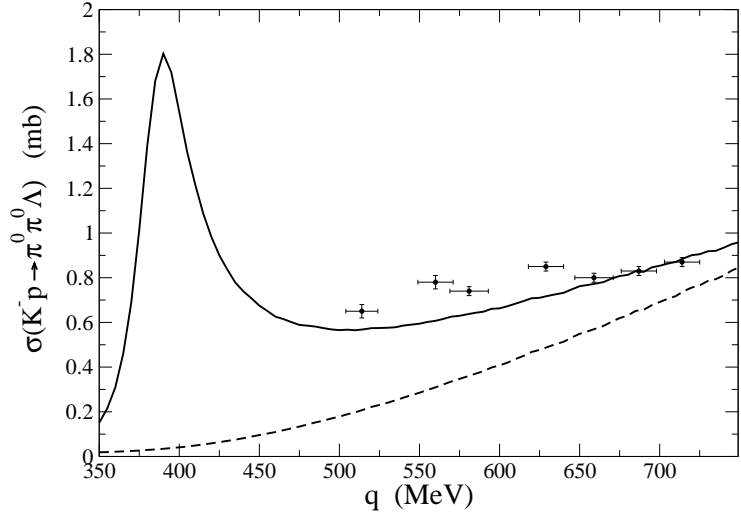}
}
\caption{Result for the $K^-p\to\pi^0\pi^0\Lambda$ cross section.
Experimental data from Ref.~\cite{Prakhov:2004ri}}
\label{fig:nefkens}
\end{figure}

\begin{figure}
\centerline{
\includegraphics[width=0.6\textwidth]{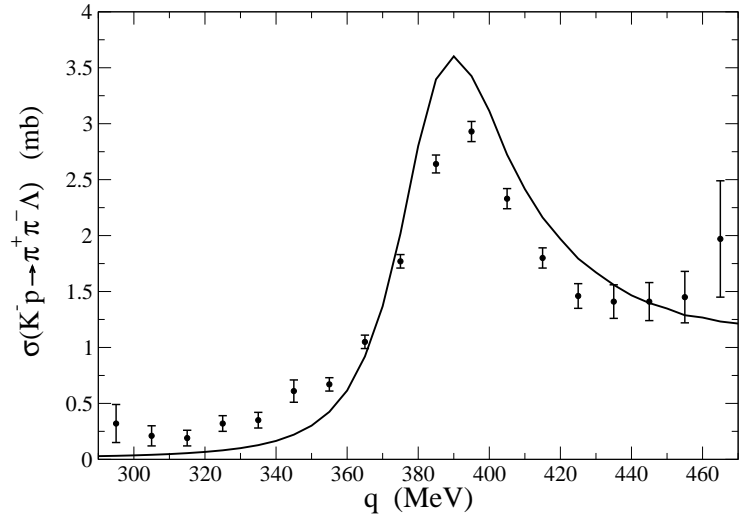}
}
\caption{Result for the $K^-p\to\pi^+\pi^-\Lambda$ cross section.
Experimental data from Ref.~\cite{Mast:1973gb}.}
\label{fig:mast}
\end{figure}

In Figs.~\ref{fig:nefkens} and \ref{fig:mast} 
 we show our results for 
$K^-p\to\pi^0\pi^0\Lambda$ and $K^-p\to\pi^+\pi^-\Lambda$
 cross section respectively along with experimental data
from Refs.~\cite{Prakhov:2004ri,Mast:1973gb}.
The dashed line represents the contribution from mechanisms 
 other than the unitarized
coupled channels, and the solid gives the coherent sum of all
the processes.
Note that the cross section of the $K^-p\to\pi^+\pi^-\Lambda$ reaction
is a factor two larger at the peak than the $K^-p\to\pi^0\pi^0\Lambda$
one.
These cross sections depend essentially on the $T_{\bar K
N\to\pi\Sgs}$ amplitude, which is obtained from our coupled channel
analysis. It is  a non-trivial prediction of the theory  since this
amplitude has not been included in the fit. Actually, the strength
at the peak comes essentially from the unitarity constraint of the
theory in analogy with the discussion above regarding the $g_1$
coupling.

\begin{figure}
\centerline{
\includegraphics[width=0.6\textwidth]{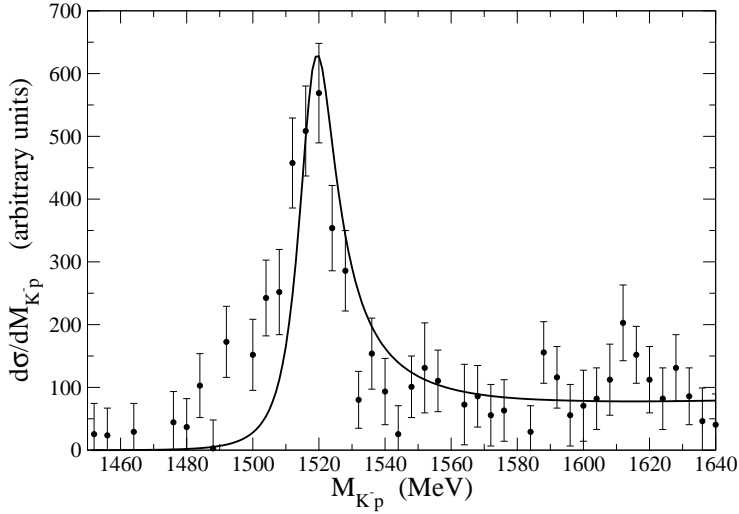}
}
\caption{$K^-p$ invariant mass distribution for the
 $\gamma p\to K^+ K^-p$ reaction with photons in the range
 $E_\gamma=2.8-4.8\textrm{ GeV}$.
Experimental data from \cite{Barber:1980zv}.}
\label{fig:dares}
\end{figure}

We now evaluate the $K^-p$ invariant mass distribution for the
$\gamma p\to K^+ K^-p$ reaction. In Ref.~\cite{Roca:2004wt} the basic
phenomenological model is explained but there only the $\pi\Sigma^*$
and $K\Xi^*$ channels were considered.
The amplitude reads now
\be
t_{\gamma p\to K^+ K^-p}=V_1\sqrt{\frac{2}{3}}k
G_{\pi\Sigma^*}
T_{\pi\Sigma^*\to\bar{K}N}\left(\frac{-1}{\sqrt{2}}\right),
\label{eq:tphoto1}
\ee
\noindent
where $k$ is the photon energy and $V_1$ the $\gamma 
p K^+\pi\Sigma^*$ vertex function which we assume to have a
smooth energy dependence.
In Eq.~(\ref{eq:tphoto1}) the last numerical factor is an isospin
coefficient to project the $K^-p$ state into $I=0$.
This amplitude is effectively written in such a way that when
taking $|t|^2$ one is automatically summing and averaging over
final and initial spin polarizations respectively.
  In Fig.~\ref{fig:dares} we
present the results obtained with the formalism of the present work
along with the experimental data of \cite{Barber:1980zv},
where we have removed $50$ of the arbitrary units in the
background which are apparent below the resonant peak.
This background could be associated to the Drell
 mechanism ($K$ exchange in the
$t$-channel), as discussed in \cite{Sibirtsev:2005ns}. 
 Given
the fact that the absolute strength of the results is a free
parameter, the results do not change qualitatively if this
background is not removed.
The normalization is arbitrary in the experimental data as well
as in our calculation. For the purpose of the present work the
shape of the distribution is the most important part and we can
see that the agreement is quite fair. It is worth mentioning
that in the work of Ref.~\cite{Roca:2004wt} the reaction was
initiated by a $\pi\Sgs$ loop. However, with the new channels
considered in the present work, initial $\bar K N$ and
$\pi\Sigma$ loops are also possible. The coupling of these
channels to $\gamma p K^+$ can be different. However we have
checked that the shape of the distribution for all these
possibilities is very similar  around the $\Ls$ peak. For higher
energies this is also true within about $30$\%. 

Note that for our purpose we do not need an explicit model for the
photoproduction of the $\Lambda(1520)$, the $V_1$ term of 
Eq.~(\ref{eq:tphoto1}). All we are stating is that there is a
doorway for the production of the $\Lambda(1520)$ through the
intermediate production of $\pi\Sigma(1385)$ (or $\bar{K}N$,
$\pi\Sigma$, states). In the literature there are explicit models
for the $\gamma N\to K \Lambda(1520)$ reaction. In one of them
\cite{Sibirtsev:2005ns} it is shown that $K$ exchange in the
$t$-channel cannot explain the data of Ref.~\cite{Barber:1980zv}
and consequently the weight of the cross section is put in $K^*$
exchange, for which the appropriate helicity amplitudes are
parametrized to reproduce the data. However, the situation, as
clearly stated in \cite{Sibirtsev:2005ns}, is that the  $\gamma
p\to K^+\Lambda(1520)$ reaction  can be more elaborate than
assumed in their model, or other models, to the point that they
could not provide a precise determination of the
$K^*N\Lambda(1520)$ coupling.  In Ref.~\cite{Titov:2005kf}, $s$-
and $u$-channels plus a contact $\gamma N K \Lambda(1520)$ term
are also introduced and a parametrization is used based on Regge
trajectories. Two solutions are given for the $N
K^*\Lambda(1520)$ coupling which, in terms of the $N
K\Lambda(1520)$ one, was given by $\alpha g_{N K\Lambda(1520)}$,
with $\alpha=0.37$ or $-0.66$, which gives us an idea of the
strength of the coupling. However, in Ref.~\cite{Nam:2005uq}, a
different parametrization of such terms is done, without using
Regge trajectories, but introducing explicitly form factors, by
means of which a good reproduction of the $\gamma p\to
K^+\Lambda(1520)$ data is obtained which would be compatible with
having a $N K^*\Lambda(1520)$ coupling equal zero, and the largest
contribution comes from the contact term. In a more recent paper
\cite{new} the $K^*$ coupling to $N\Lambda(1520)$ is evaluated
using the present model and a simple quark model for comparison.
The coupling obtained in \cite{new}
with the present model is smaller than in the quark
model and also than in \cite{Titov:2005kf}. As we can see from the
previous discussion, the role of the $K^*$ exchange mechanism in
the $\gamma p\to K^+\Lambda(1520)$ reaction is rather
controversial. However, for the purpose of the present work this
information is not needed. It would be absorbed in  our $V_1$
term of Eq.~(\ref{eq:tphoto1}) which is unnecessary in our
comparison with the 
unnormalized cross section, or in ratios of cross
sections as we discuss below.

In Ref.~\cite{Roca:2004wt} it was suggested that the
ratio of the $\pi^0\pi^0\Lambda$
mass distribution of the 
$\gamma p\to K^+\pi^0\pi^0\Lambda$ to the $K^-p$ distribution of the 
$\gamma p\to K^+ K^-p$ reaction at the peak of the
 $\Ls$ resonance
can  provide a test of the coupling of the $\Ls$ to
$\pi\Sigma^*$ ($g_1$). 
In an analogous way to  the $K^-p\to\pi^0\pi^0\Lambda$ reaction,
the $\gamma p\to K^+\pi^0\pi^0\Lambda$ would proceed through 
$\gamma p\to K^+\pi^0(\Sigma^*)\to K^+\pi^0(\pi^0\Lambda)$.
The amplitude is 
\ba \nonumber
t_{\gamma p\to K^+\pi^0\pi^0\Lambda}(p1,p2)&=&
V_1\sqrt{\frac{2}{3}}kG_{\pi\Sigma^*}
T_{\pi\Sigma^*\to\pi\Sigma^*}\frac{-1}{\sqrt{3}}\,\cdot \\
&\cdot&\frac{1}
{\sqrt{s_{\Sgs}}-M_{\Sgs}+i\Gamma_{\Sgs}(\sqrt{s_{\Sgs}})/2}
\frac{f_{\Sgs\pi\Lambda}}{m_\pi}
\frac{1}{\sqrt{2}}(p_{2x}-ip_{2y}),
\ea
where one has to consider the same symmetrization arguments as in
the $K^-p\to\pi^0\pi^0\Lambda$ reaction
and $\vec p_2$ is taken in the rest frame of the $\Sgs$.
Once again this amplitude is effectively written in such a way that when
taking $|t|^2$ one is automatically summing and averaging over
final and initial spin polarizations respectively.
Up to phase space and the rest of numerical factors
the ratio
$R=(d\sigma_{\gamma p\to K^+\pi^0\pi^0\Lambda}
/dM_{\pi^0\pi^0\Lambda}) /
(d\sigma_{\gamma p\to K^+K^-p}
/dM_{K^-p})$
is proportional to $(|T_{\pi\Sigma^*\to\pi\Sigma^*}|
/|T_{\pi\Sigma^*\to\bar{K}N}|)^2$ 
which, at the $\Ls$ peak position, is $(g_1/g_3)^2$
(see Eq.~(\ref{eq:Tcoup})).
All considered we obtain a value for the ratio of  $R\sim 0.25\pm
0.04$,
while $(g_1/g_3)^2$ is $2.8$,  where the error of $R$ 
has been obtained
propagating the errors of Table~\ref{tab:coup}.

In Fig~\ref{fig:dares} we see that our model seems to be lacking some strength at
energies below the $\Lambda(1520)$ peak. This could be due to lack of some
background terms or the large uncertainty of the experimental errors. In fact, it
is worth stressing that the same feature appears in the result of
Ref.~\cite{Sibirtsev:2005ns}. On the other hand, more recent experimental results
\cite{Barth} and \cite{baltzel}, although at different photon energies, do not show this broad
shoulder.\\

\begin{figure}
\centerline{
\includegraphics[width=0.8\textwidth,angle=0]{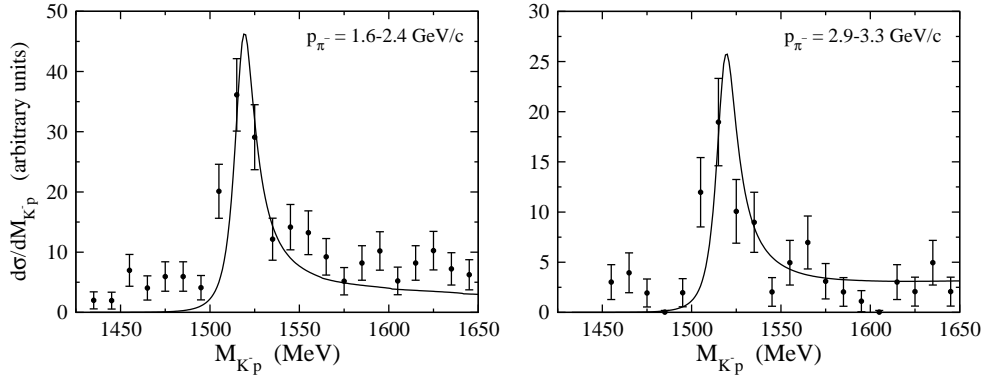}
}
\caption{$K^-p$ invariant mass distribution for the
 $\pi^- p\to K^0 K^-p$ reaction. Left plot: pions in the range
 $p_{\pi^-}=1.6-2.4\textrm{ GeV/c}$; right plot:
 $p_{\pi^-}=2.9-3.3\textrm{ GeV/c}$.
Experimental data from Ref.~\cite{dahl}.}
\label{fig:dahl}
\end{figure}

Next we show, in Fig.~\ref{fig:dahl},
the $K^-p$ invariant mass distribution for
 the $\pi^- p\to K^0 K^-p$ reaction. The model used 
 for this purpose is analogous
to the one used in the study of the 
$\gamma p\to K^+K^-p$ reaction. 
The differences are in the phase space and in the first vertex where
we have used a different unknown constant.
In  Ref.~\cite{dahl} no errors in the experimental data are provided,
hence we have assumed errors to be the square root of the number of
events (statistical errors).
We see that the agreement is fair considering the large experimental
errors. 
Note that, unlike in the case of photoproduction, we did not subtract a background
here, which would clearly make the agreement
with experiment better at energies below
 the $\Lambda(1520)$ region.

\section{Conclusions}

We have done a coupled channel analysis of the $\Ls$ resonance
using the  $\pi\Sigma(1385)$, $K\Xi(1530)$, $\bar{K}N$ and
$\pi\Sigma$ channels. We have used the Bethe-Salpeter equation
to implement unitarity  in the evaluation of the different 
amplitudes. The main novelty from previous coupled channel
approaches to this resonance is the inclusion of new matrix
elements  in the kernel of the BS equation and the consideration
of the $\Sigma(1385)$ width in the $\pi\Sigma^*$  loop function.
The unknown parameters in the $V-$matrix, as well as the
subtraction constants of the loop functions, have been obtained
by a fit to $\bar K N\to\bar K N$ and $\bar K N\to \pi\Sigma$
partial wave amplitudes. As a consequence of the unitarity of
the scheme used, we can predict the amplitudes and couplings of
the $\Ls$ for all the different channels.  The largest coupling
is obtained for the $\pi\Sgs$ channel.

We have then tested the amplitudes obtained in several specific
reactions and compared with experimental data at energies 
close to and slightly above the $\Ls$ region. These include the
$K^-p\to\Lambda\pi\pi$, 
$\gamma p\to K^+K^-p$,
$\gamma p\to K^+\pi^0\pi^0\Lambda$
 and $\pi^- p\to K^0 K^-p$ reactions.
We have obtained a reasonable agreement with the 
experimental results
that allows us to be confident in the procedure followed
 to describe the
nature of the $\Ls$ resonance.

For the $K^-p\to\Lambda\pi\pi$ process, both with neutral and
charged pions, we could do predictions of the absolute cross
sections since the process involves the amplitude for the $K^-p
\to \pi \Sigma^*$ transition, which is an output of our theory. 
In the other two cases the strength of the cross section  was
left as a free parameter since no theory was done for the
coupling of the photon or the pion to the $\Lambda(1520)$
components. Yet, it was found that in all cases the shape of the
resonance was rather asymmetric, with a substantial strength
above the peak of the resonance in all the reactions, which
seems to be an intrinsic property of the $\Lambda(1520)$
resonance, and which in our theoretical framework could be
attributed to the $d$-wave character of the $\bar K N$ and 
$\pi\Sigma$ channels, as well as to the fact that above the
resonance peak the phase space for the $\Lambda(1520)$ decay
into the $\pi \Sigma^*$ channel grows rapidly. 

    We could see that, in spite of the relevant role played by
the $\bar K N$  and  $\pi\Sigma$ channels, the $\Lambda(1520)$
remains with the largest coupling to the $\pi \Sigma^*$ channel.
In a simplified theory in which only the  $s$-wave $\pi
\Sigma^*$ and $ K \Xi^*$ channels are taken, and their
interaction is provided by the chiral Lagrangians,  the
$\Lambda(1520)$ appears naturally and qualifies as a quasibound
$\pi \Sigma^*$ state.  The coupling of the resonance to the
$\bar K N$ and $\pi\Sigma$ channels in the real world modifies
appreciably this picture but the state still keeps memory of the
original simplified picture and has still a large coupling to
$\pi \Sigma^*$ with a strength of $g^2$ of the order of three times
larger than that of the  $\bar K N$ or $\pi\Sigma$ channels.

    Further tests to show the dominance of the $\pi \Sigma^*$ component in the 
$\Lambda(1520)$ resonance could be done, in particular a crucial test would be
the modification of the $\Lambda(1520)$ width in nuclei, and particularly the 
$\pi \Sigma^*$ mode that in nuclei would give rise to the $ph \Sigma^*$ decay
channel.  In this way one could observe a drastically enhanced production of 
$\Sigma^*$ in processes involving $\Lambda(1520)$ decay in nuclei, while at the
same time one would observe an increased $\Lambda(1520)$ width in other physical
processes.  The  understanding of the nature of this resonance and the
comprehension of processes where the resonance appears would benefit much from
the performance of such experiments. 

\section*{Acknowledgments}
This work is partly supported by DGICYT contract number BFM2003-00856,
and the E.U. EURIDICE network contract no. HPRN-CT-2002-00311.
This research is part of the EU Integrated Infrastructure Initiative
Hadron Physics Project under contract number RII3-CT-2004-506078.

\end{document}